International Journal of Engineering Trends and Technology (IJETT) – Volume 13 Number 2 – Jul 2014# Data Transfer between Two USB Flash SCSI Disks using a Touch Screen

Anurag A. Chakravorty[#1], Raghwendra J. Suryawanshi[*2],

[#] Bachelor of Engineering, Department of Information Technology,

Matsyodari Shikshan Sanstha's College Of Engineering & Technology, Jalna,

(Dr. Babasaheb Ambedkar Marathwada University,

Aurangabad, Maharashtra, India)*Abstract—* In this information age, importance of data in human life is no more just secondary. Electronics plays a vital role in our daily life. In the present world of electronics, there are various ways that are present for storage of any type of data electronically, today's most used & portable device is pen drives or USB mass storage devices.

Under normal circumstances, as an intermediate device, if we want to move or copy data from one mass storage device to another, we use a computer in the form of desktops, laptops, etc. We need a device which can be used as an intermediate device, also which is a complete blend of hardware & software. This device is a gadget that can be used to transfer data between two flash SCSI devices via a touch screen.

This is a user friendly device which uses the most popular bus – USB (Universal Serial Bus) with Type-A connector. It is governed by the USB 2.0 Protocol. One of the major advantage of this device is its portability.

*Keywords—* **USB Mass Storage drives, FAT, SCSI, VNC2, FT800, Pen drive.**## I. INTRODUCTION

The USB to USB Bridge can be used an intermediate device to transfer data between two USB flash SCSI devices instead of using computer as an intermediate for the same purpose. Carrying a laptop or desktop for only data transfer purpose is not affordable these days when people want all devices to be handy. Moreover, transferring data using a computer involves a lot of power to be wasted as we first have to wait for it to boot up, then plug in the device & then transfer the data. So, a lot of time is wasted.

Also, the threat of viruses & malware has made the life of computer users more complicated. These viruses get activated as soon as the device is plugged into the system. This device (the USB to USB Bridge) is perfect to eliminate these disadvantages.

This device will have the following features:-

- Handy, portable & lightweight device.
- Supports all USB storage SCSI devices in FAT 16 & 32 formats.
- Device can be powered using 9V Battery.
- Supports USB 2.0.
- Plug & play feature.

### A. USB

USB stands for Universal Serial Bus. **Universal Serial Bus** (**USB**) is an industry standard that defines the cables, connectors and communications protocols used in a bus for connection, communication, and power supply between computers and electronic devices. USB was designed to standardize the connection of computer peripherals (including keyboards, pointing devices, digital cameras, printers, portable media players, disk drives and network adapters) to personal computers, both to communicate and to supply electric power.

USB has effectively replaced a variety of earlier interfaces, such as serial and parallel ports, as well as separate power chargers for portable devices.

We will be using Type-A standard USB 2.0 connector. The standard pin description along with standard cable color is displayed in table 1

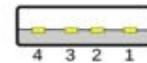

Fig. 1 USB Type-A Connector[9]

TABLE I
USB TYPE-A CONNECTION

| USB 2.0 standard pin-out | | | |
|---|---|---|---|
| Pin | Name | Cable colour | Description |
| 1 | VBUS | Red (or Orange) | +5 V |
| 2 | D− | White (or Gold) | Data − |
| 3 | D+ | Green | Data + |
| 4 | GND | Black (or Blue) | Ground |

### B. Mass Storage Detection Basics

On power-up, hubs make the host aware of all the attached USB devices. In a process called **enumeration**, the host assigns an address & requests a series of data structures called **descriptors** from each device. After power-up, whenever a device is removed or attached, the host learns of the event & enumerates any newly attached device or removes any detached device from the record of available devices.

ISSN: 2231-5381            http://www.ijettjournal.org            Page 71

## II. REQUIREMENTS OF THE SYSTEM

The concept of this device was invented so as to answer the following question; what can be the solution to make the computer independent data transfer via USB? The only available answer resulted in the development of a system that will perform the same task as that of a computer. Making a system that will do the tasks of a computer – made us to think a little because the answer directly suggests implementing a system that will handle USB protocol along with the other processes.

*Requirement 1*: The first & foremost need is that there should be a device which supports USB 2.0 protocol.

*Requirement 2*: The speed of transfer is to be taken under consideration which should be equal to the full & high speed USB compatible.

## III. IMPLEMENTATION

### A. HARDWAREs:

The selection of hardware is an important task in developing this device. The main hardware is the processor or the microcontroller which will drive the unit. The secondary hardware needed is the video graphics controller which will take care of the RGB colour scheme, touch screen & the audio using PWM signals. Here, the touch screen is used to provide menu driven system that will put the options on screen.

*1) Vinculum-II:* This is the main microcontroller used in the development of this system. The FTDI Vinculum-II (VNC2) Embedded Dual USB Host Controller IC represents the second generation of FTDI USB Host devices. The Vinculum-II CPU is dramatically upgraded from the previous VNC1L device, greatly increasing the processing power. This FTDI Embedded Dual USB Host Controller features a powerful 16-bit MCU core with 256KB Flash and 16KB RAM memory. The Vinculum-II is a complete USB system solution that supports a range of flexible interfaces, including UART, SPI, FIFO, and PWM. The complete USB protocol data processing is handled entirely by hardware resources within the device, freeing up processing resources for user developed applications.

*VNC2 Advanced Features*
- Embedded processor core.
- 16 bit Harvard architecture.
- Two full-speed or low-speed USB 2.0 interfaces capable of host or slave functions.
- 256Kbytes on-chip E-Flash Memory (128K x 16-bits).
- 16Kbytes on-chip Data RAM (4K x 32-bits).
- Programmable UART up to 3Mbaud.
- Two SPI (Serial Peripheral) slave interfaces and one SPI master interface.
- Reduced power modes capability.
- Variable instruction length.
- Native support for 8, 16 and 32 bit data types.
- Eight bit wide FIFO Interface.
- Firmware upgrades via UART, SPI, FIFO interface or USB Flash Drive.
- 12MHz oscillator using external crystal.
- General-purpose timers.
- Software development suite of tools to create customized firmware.
- Compiler Linker - Debugger - IDE.
- 44 configurable I/O pins on the 64 pin device, 28 I/O pins on the 48 pin device and 12 I/O on the 32 pin device using the I/O multiplexer.
- +3.3 volt supply.
- -40°C to +85°C extended operating temperature range.
- Simultaneous multiple file access on BOMS devices.
- Eight Pulse Width Modulation outputs to allow connectivity with motor control applications.
- Debugger interface module.
- System Suspend Modes.

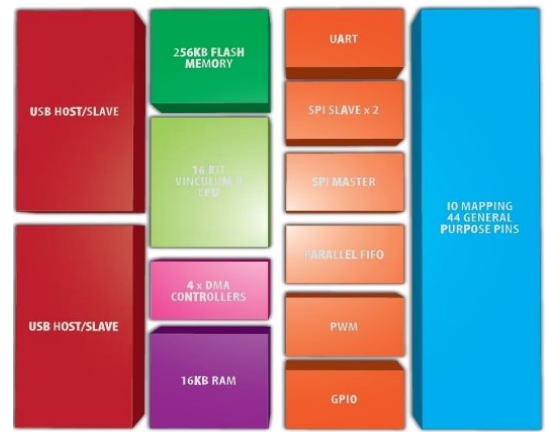

Fig. 2 Block Diagram of Vinculum-II[2]

*2) FT-800(EVE):* EVE is FTDI's series of Human Machine Interface (HMI) controller ICs**. FTDI** Embedded Video Engine (EVE) features an initial chip offering of the FT800, which addresses the need for easy-to-design, advanced forms of human-machine interaction. EVE provides a high quality graphic solution with 3-in-1 functionality for graphical user interface (GUI) development. The FT800 combines display, audio, and touch operations into a single chip, providing an optimized solution that reduces power, board area, BOM costs, and much more.

With these operations, engineers now have an advanced solution to easily create and output state-of-the-art interactive display systems. Targeted at intelligent QVGA and WQVGA TFT display panels, EVE's object-oriented approach renders images in a line-by-line fashion with 1/16th of a pixel resolution, eliminating the expense of traditional frame buffer memory. The FT800 integrates with system MCUs via a low bandwidth serial interface, allowing for lower spec/cost MCUs to be used in the design.

The controller's functionality sets new industry benchmarks, supporting 4-wire resistive touch sensing with built-in intelligent touch detection and an embedded audio processor that allows midi-like sounds, combined with pulse code modulation (PCM) for audio playback. The combination of display, audio, and touch on a single-chip solution enables engineers to produce GUIs that deliver compelling user experiences.

- Support for inline assembly
- Efficient RAM usage and optimizations
- Separate preprocessor (VinCpp.exe)
- Produces optimized code for VNC2

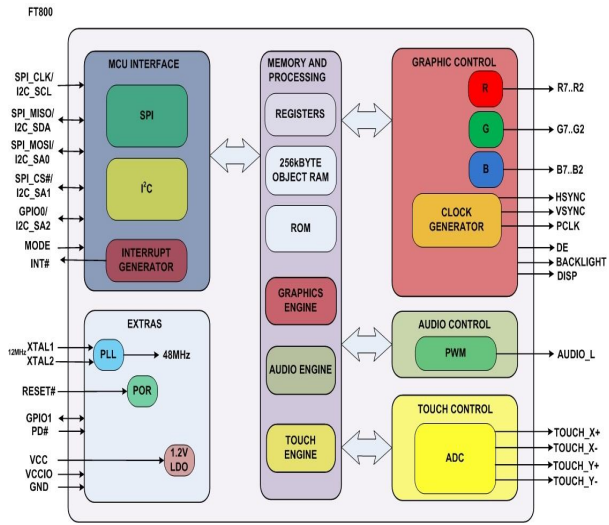

Fig. 3 Block Diagram of FT800 (EVE) *[3]*

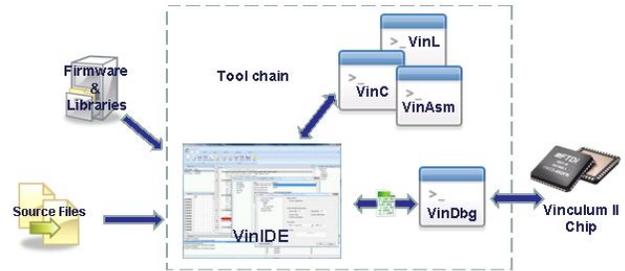

Fig. 4 Vinculum II IDE*[1]*

*3) LCD & Touch screen module:* Any intelligent QVGA or WQVGA TFT display panel that is compatible with FT800 can be used. A compatible touch panel module is required.

*4) Speakers:* Speakers supporting PWM (Pulse Width Modulation)   compatible with FT800 is to be used.

B. *Software's:*

 *1) Vinculum-II IDE*: FTDI has created set of tools for Vinculum II (VNC2) which includes a C compiler, assembler, linker, debugger and integrated development environment. These tools facilitate application development on VNC2 using a kernel, device driver and runtime libraries provided by FTDI.

All tools are command line applications and as such can be integrated into third party applications such as IDEs or scripts. VinC Compiler, implemented as part of the overall Toolchain, for the VNC2 is:

- ANSI 'C' compatible (with restrictions)
- Support for structures, unions and arrays - Structures and arrays can be comprised of base data types or other data structures
- Language level support for accessing flash memory
- Support for pointers including function pointers.
- Support for typedef().
- There are some restrictions on using pointers to data stored in ROM.
- Support for ANSI C control flow statements, selection statements and operations

*2) Windows Operating System:*  A Windows based operating system is required to run Vinculum-II IDE & all its other components.

IV. ARCHITECTURE OF THE SYSTEM

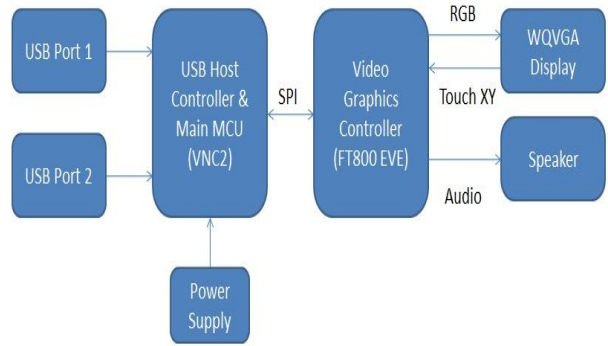

Fig. 5 System Block Diagram

As shown in figure 5, the architecture of the system mainly consists of a two processors viz., USB Host controller & video graphics controller. The USB Host Controller also acts as the main microcontroller taking care of the system's operation. Thus, here we eliminate the need for two processors & use instead only one processor acting as both USB Host Controller & main MCU.

This processor is the heart of the device. The flexibility & choices in interfaces as well as robustness provided by the VNC-2 controller is very high.

This microcontroller meets the requirement of a processor that can handle fast transfers of data. Now the next microcontroller in the architecture is FT800 also popularly known as EVE (Embedded Video Engine). It handles the entire display, touch as well as audio features.

The first one is the Vinculum 2 microcontroller while the latter one is the FT800 EVE controller handling touch, display as well as monotonic audio.

It is interfaced to VNC-2 via SPI (Serial Peripheral Interface) interface. The **Serial Peripheral Interface Bus** or **SPI** bus is a synchronous serial data link *de facto* standard, named by Motorola that operates in full duplex mode. Devices communicate in master/slave mode where the master device initiates the data frame.

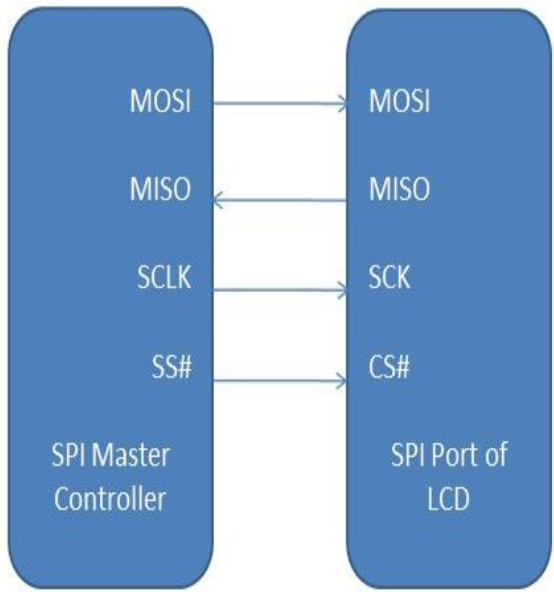

Fig. 6 SPI Master Controller to SPI Port of LCD[8]

Various commands are used to access the information such as files, folders, total size & other system related information of the USB device. Once the USB mass storage device is connected, the contents of the pen drive are displayed on the LCD.

Note: Only SCSI (Small Computer System Interface) is supported & ATAPI (ATA Parallel Interface) devices will be supported in the future upgraded versions.

## V. WORKING OF THE SYSTEM

### A. System flow:

The connectivity type here is peer-to-peer communication. USB host controller is the main microcontroller used to handle both the touch LCD as well as both the flash drives. The USB host controller uses the File System Controller which actually use retrieve commands to extract the necessary data from the main USB host controller.

The File System Controller is the intermediate controller (logically) between the USB Host controller & the Video user interface controller which handles the System memory as well The first one is the Vinculum 2 microcontroller while the latter one is the FT800 EVE controller handling touch, display as well as monotonic audio.

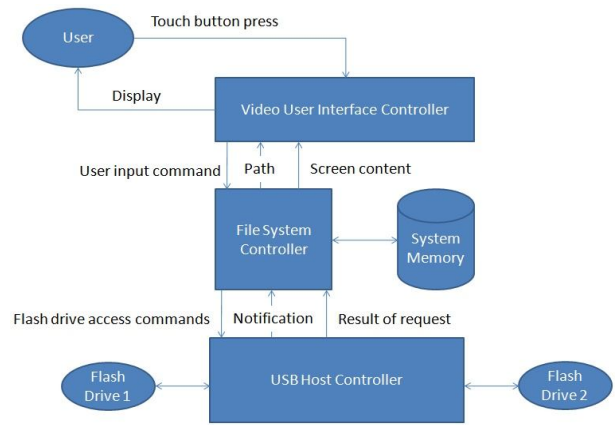

Fig. 7 System Flow Diagram

### B. Implementation for experiment purposes

For implementation purposes we have the VNC2 based V2EVAL board i.e. it is an evaluation board for VNC2 to carry out ours experiments. We have found successful results of our theory & the data can thus be transferred from one USB to the other USB using a touch screen.

## VI. ADVANTAGES

*1) Portability:* Making the proposed system a standalone system, the device can be used as a portable device. The device works i.e. carries out transfers independently without using computer systems. So, it can out transfers anywhere, anytime.

*2) Battery operated & power optimization:* We can power-up the system using a 9 volt battery thus supporting portability. Both the processor & USB 2.0 specifications are designed for power optimization.

## VII. CONCLUSIONS

The device gives us an opportunity to understand the USB 2.0 protocol & develop a device that is USB 2.0 compatible.

A set of basic requirements were defined & used for design work of the USB concept based on practical tests & results. We also used only one microcontroller instead of separate microcontrollers for USB Host Controller & managing the touch TFT panel.

The resulting device is a portable device which is battery operated & can be used to transfer data anywhere, anytime.

## VIII. FUTURE SCOPE

While working for the development of this device and exploring the peripherals that can be interfaced with the VNC-2 we found that with a little modification, we can add the following features to the device:

- Add support for ATAPI peripheral devices using the USB97C202 IC.

- Support for NTFS flash drives.
- Using Bluetooth we can also connect to other Bluetooth devices thus supporting wireless transfers.
- Using GPS, anti-theft tracker can also be implemented.
- USB Music media playback interface can also be introduced using proper integrated circuit for mp3 decoding.
- Increasing the speed of USB data transfer implementing advanced USB standards like USB 3.0 & USB 3.1 with Cypress Semiconductor chips.
- Capacitive touch LCD.